\documentclass[12pt,eps]{revtex4}
\usepackage{amssymb}
\usepackage{epsfig}

\def\beq{\begin{equation}}
\def\eeq{\end{equation}}
\def\beqa{\begin{eqnarray}}
\def\eeqa{\end{eqnarray}}
\def\hf{\textstyle{1\over2}}
\def\3hf{\textstyle{\frac{3}{2}}}

\newcommand{\ket}[1]{\vert #1 \rangle}
\newcommand{\cg}[6]{C^{#5\,#6}_{#1 , #2 ; #3 , #4}}
\newcommand{\bra}[1]{\langle #1 \vert}
\newcommand\unit{\mathinner{\hbox{1}\mkern-4mu\hbox{l}}}

\newcommand{\binom}[2]{ {#1 \choose #2}}
\newcommand{\e}{\hbox{\rm e}}
\newcommand{\myfrac}[2]{\leavevmode\kern.1em\raise.5ex\hbox{\scriptsize
$#1$}\kern-.1em {\scriptsize
/}\kern-0.10em\lower.25ex\hbox{\scriptsize $#2$}}

\begin{document}

\title[su(2) intelligent states]{Su(2) intelligent states as coupled su(2) coherent states}

\author{Benjamin R. Lavoie and Hubert de Guise}
\address{Department of Physics, Lakehead University,
Thunder Bay, ON, P7B 5E1, Canada}


\begin{abstract}
We show how su(2) intelligent states can be obtained by coupling
su(2) coherent states. The construction is simple and efficient, and
easily leads to a discussion of some general properties of su(2)
intelligent states.
\end{abstract}

\date{\today}
\maketitle

\section{Introduction}

In quantum mechanics, uncertainty relations give a lower bound on
the uncertainty resulting from the simultaneous measurement of two
non--commuting observables. One common uncertainty relation was
obtained in
\cite{robertson}: if $\hat \Omega$ and $\hat \Lambda$ are
self-adjoint operators, and if $\left\vert \psi \right\rangle $ is a
state normalized to $1$, then we have
\begin{equation}
\Delta \Omega\Delta \Lambda\geq \textstyle\frac{1}{2}\,\vert
\langle[ \hat \Omega,\hat \Lambda ] \rangle \vert .
\label{robertson}
\end{equation}%
In Eq.(\ref{robertson}), $\Delta \Omega$ is the standard deviation
of the operator $\hat \Omega$ for a quantum system described by
$\left\vert \psi
\right\rangle ,$ \textit{i.e.}%
\begin{equation}
\Delta \Omega=\sqrt{\langle \hat \Omega^{2}\rangle -\langle\, \hat
\Omega\,\rangle ^{2}},
\end{equation}%
with $\langle\,\hat X \rangle =\langle \psi \vert\, \hat X \, \vert
\psi \rangle .$

In this paper, we will discuss su(2) states for which the strict
equality in Eq.(\ref{robertson}) holds, \textit{i.e.} su(2) states
for which ($\hbar =1$)
\begin{equation}
\Delta L_{x}\Delta L_{y}=\textstyle\frac{1}{2}\vert \langle \hat{L}
_{z}\rangle\vert .  \label{intelligentdef}
\end{equation}

States that satisfy Eq.(\ref{intelligentdef}) are known as su(2)
intelligent states. The terminology was first introduced by Aragone
\textit{et al} \cite{aragone}. It is clear that the right hand side
of Eq.(\ref{intelligentdef}) depends on the choice of state used to
evaluate $\langle \hat{L} _{z}\rangle$, so intelligent states need
to be distinguished from minimum uncertainty states; there are
intelligent states for which the rhs of Eq.(\ref{intelligentdef}) is
not the obvious minimum value of
$0$.

By su(2) state, we understand a (pure) quantum state $\ket{\psi}$
that belongs to an irreducible representation of the su(2) algebra.
This algebra is spanned by the familiar angular momentum operators
$\{\hat L_x,\hat L_y,\hat L_z\}$ or, more conveniently, by the
complex linear combinations $\{\hat L_{\pm},\hat L_z\}$, where
\beq
\begin{array}{rclcccc}
\hat L_\pm                        &=&\hat L_x\pm i \hat L_y\, , &&&& \\
\left[\hat L_z,\hat L_{\pm}\right]&=&\pm\, \hat L_{\pm}\, , &\quad&
\left[\hat L_+,\hat L_-\right]&=&2\hat L_z\, .
\end{array}
\eeq
An irreducible representation of dimension $2j+1$, where $j$ can be
an integer or a half--integer, is spanned by the set
$\{\ket{jm},m=-j,-j+1,\ldots,j-1,j\}$ with
\beq
\hat L_z\ket{jm}=m\,\ket{jm}\, ,\qquad \hat
L_{\pm}\ket{jm}=\sqrt{(j\mp m)(j\pm m+1)}\ket{j,m\pm 1}\, .
\eeq


Intelligence is not limited to su(2) states. A well--known example
of intelligent states is the harmonic oscillator coherent state
$\left\vert\, \xi \right\rangle $, parameterized by the complex
number $\xi$ and for which
\begin{equation}
\Delta x\Delta p=\textstyle\frac{1}{2}\, .
\end{equation}
However, in this paper, we understand intelligent states to mean
su(2) intelligent states.

The terminology ``su(2) intelligent states'' is to be contrasted
with recent theoretical and experimental work
\cite{FrankeArnold}\cite{Pegg}\cite{hradil} on angular momentum
states of light as quantum states carrying orbital angular momentum
about the beam axis.  In these papers, the spectrum of the operator
$\hat L_z$ is unbounded, leading to a differential
eigenvalue equation rather than the finite--dimensional eigenvalue
problem of Eq.(\ref{intelligenceeqn}).

An important ingredient to our construction will be the su(2)
coherent states \cite{su2coherent}. It is sufficient here to recall
the well--known property that such states are obtained by a rotation
of the extremal su(2) state $\ket{\ell,\ell}$. More specifically, an
su(2) coherent state $\ket{\gamma,\vartheta}$ can be parameterized
by two angles $\gamma,\vartheta$ such that, up to an overall phase
\beq
\ket{\gamma,\vartheta}=R_z(\gamma)R_y(\vartheta)R_z(-\gamma)\ket{\ell\,
,\ell}\, , \label{su2coherent}
\eeq
where $R_i(\varphi)$ denotes the rotation about the axis $i$ by an
angle $\varphi$. Su(2) coherent states with $\gamma=0$
 or $\pi/2$
also satisfy
 Eq.(\ref{intelligentdef}) . However, su(2)
intelligent states are not always of the form of
Eq.(\ref{su2coherent}).

Indeed, we plan to show that all intelligent states are of the
form
\beqa
&&\left[R_y(\beta)\ket{\ell_A\, ,\ell_A}\right]\otimes
\left[R_y(-\beta)\ket{\ell_B\, ,\ell_B}\right]\, ,\\
\hbox{\rm or }&& \left[R_x(\beta)\ket{\ell_A\,
,\ell_A}\right]\otimes \left[R_x(-\beta)\ket{\ell_B\,
,\ell_B}\right]\, ,
\eeqa
corresponding to Eqn.(\ref{su2coherent}) with $\gamma=0$ or $\pi/2$
and a specific choice of $\vartheta$.

Su(2) intelligent states of angular momentum $\ell$ are of the form
\beqa
&&\hat\Pi^{\ell}\, \left[R_y(\beta)\ket{\ell_A\,
,\ell_A}\right]\otimes \left[R_y(-\beta)\ket{\ell_B\,
,\ell_B}\right]\, , \label{introprojected}\\
\hbox{\rm or }&&\hat\Pi^{\ell}\, \left[R_x(\beta)\ket{\ell_A\,
,\ell_A}\right]\otimes \left[R_x(-\beta)\ket{\ell_B\,
,\ell_B}\right]\, ,\nonumber
\eeqa
with $\ell=\ell_A+\ell_B$ and where
$\hat\Pi^{\ell}=\sum_m\,\ket{\ell,m}\bra{\ell,m}$ is the
(non--unitary) operator that projects into the $\ell$ subspace.

Thus, our work functions as a bridge between the work of Hillery and
Mlodinow \cite{Hillery} and the work of Rashid\cite{rashid}. In
\cite{Hillery}, some intelligent states were obtained as su(2)
coherent states. They correspond to setting $\ell_B=0$ in
Eqn.(\ref{introprojected}). No projection is required and, although
not every su(2) intelligent state can be constructed, the use of a
single unitary transformation means that these states are amenable
to experimental implementation \cite{opticalimplementation}. The
construction method of \cite{rashid} is distinctive in that it
requires the use of a non--unitary transformation, although it
completely solves the construction problem in a single shot.

Eqn.(\ref{introprojected}), on the other hand, lends itself to a
clear physical interpretation: to construct a general intelligent
state of angular momentum $\ell$, we must bring together two
separate systems, each of which has been subjected to a different
unitary transformation, and then extract from this combined system
states of good angular momentum using a non--unitary operation akin
to a measurement of $\ell$. This interpretation provides a much
clearer picture of su(2) intelligent states than the one presented
in \cite{polynomialstates}.

In addition to \cite{Hillery} and \cite{rashid}, the original work
\cite{aragone} of Aragone \emph{et al.} has blossomed in various
directions. In particular, the recent work of \cite{Nha} deals with
entanglement and su(2) intelligent states.  Generalized intelligent
states, which satisfy
\begin{equation}
\Delta \Omega^{2}\Delta \Lambda^{2}=
\textstyle\frac{1}{4}\langle \lbrack \hat{\Omega},\hat{\Lambda}%
]\rangle ^{2}+\textstyle\frac{1}{4}\langle
\{\hat{\Omega}-\left\langle \Omega\right\rangle
,\hat{\Lambda}-\left\langle \Lambda\right\rangle \}\rangle ^{2},
\label{generalrobert}
\end{equation}%
where
$\{\hat{\Omega},\hat{\Lambda}\}\equiv\hat{\Omega}\hat{\Lambda}+\hat{\Lambda}\hat{\Omega},$
have been the object of considerable attention (see, for instance,
\cite{brif1}),
including various applications in quantum optics \cite{brif2}\cite{campos}%
\cite{perinova}. Several authors, in particular \cite{agarwalpuriI},
have studied spin squeezing using the construction of \cite{rashid}.
Trifonov \cite{trifonov} has studied multi--observables and
multidimensional generalizations of Eq.(\ref{generalrobert}).

Our work is organized as follows.  We first identify a simple but
basic property of solutions of the eigenvalue problem; this is
encapsulated in Eq.(\ref{compointelligent}).  Once this is done, the
eigenvalue problem associated with intelligence is solved explicitly
for spin-$\hf$ in Sec.\ref{spinhalfcase}.  These spin-$\hf$ states
and Eq.(\ref{compointelligent}) are used in Sec.\ref{spin5halfcase}
to construct, using a minimum amount of extra work, all intelligent
states of angular momentum $\ell=\myfrac{5}{2}$. This method is
generalized to arbitrary $\ell$ in Sec.\ref{generalcase}.  The
general expression for our angular momentum state can be found in
Eq.(\ref{generalstate}). Some simple analytical and numerical
results are presented in Sec.\ref{results}. A discussion and a short
conclusion can be found in Sec.\ref{discussion}.

\section{Some simple properties}

Recall \cite{eigenvalueequation} that intelligent states $\left\vert
\psi^{\ell} \left( \alpha \right) \right\rangle $ of angular
momentum $\ell$ are
eigenstates  of the non-hermitian operator $\hat{L}_{x}-i\alpha \hat{%
L}_{y}, $ \textit{i.e.} they satisfy%
\begin{equation}
( \hat{L}_{x}-i\alpha \hat{L}_{y})\vert \psi^\ell ( \alpha ) \rangle
=\lambda\,\vert \psi^\ell ( \alpha )\rangle ,
\label{intelligenceeqn}
\end{equation}%
where $-\infty \leq \alpha \leq \infty $ is a real parameter. \ The
eigenvalue $\lambda $ is related to the average value of $\hat{L}_{x}$ and $%
\hat{L}_{y}$ and to the parameter $\alpha$ via:
\beq
\lambda=\langle\,\hat L_x\rangle-i\alpha \langle\,\hat L_y\rangle\,
.
\eeq

Equation (\ref{intelligenceeqn}) stems from two requirements. To
replace the inequality of Eq.(\ref{robertson}) by the equality and
obtain Eq.(\ref{generalrobert}), the states $( \hat{L}_{x}-\langle
\hat{L}_{x}\rangle)\,\vert \psi^{\ell} (\alpha) \rangle$ and $(
\hat{L}_{y}-\langle \hat{L}_{y}\rangle )\,\vert \psi^{\ell} (\alpha)
\rangle $ must be collinear, \textit{i.e.}
\begin{equation}
( \hat{L}_{x}-\langle \hat{L}_{x}\rangle )\, \vert \psi^{\ell}
(\alpha) \rangle =i\alpha ( \hat{L}_{y}-\langle \hat{L}_{y}\rangle
)\, \vert \psi^{\ell} (\alpha) \rangle . \label{collinear}
\end{equation}%
We obtain intelligence by forcing the anticommutator term in
Eq.(\ref{generalrobert}) to
$0$:
\beq
\bra{\psi^{\ell} (\alpha)}\{ \hat L_x-\langle \hat L_x\rangle,\hat
L_y -\langle \hat L_y\rangle \} \ket{\psi^{\ell} (\alpha)} =0.
\label{anticondition}
\eeq
This restricts the values of $\alpha $ to be real and produces
Eq.(\ref{intelligenceeqn}).

Let us now abstractly consider a composite system made from two
independent subsystems, denoted by the subscripts $A$ and $B$
respectively, such that
\beqa
\hat L_{x,A}&\equiv & \hat L_{x}\otimes \unit_B\, ,\quad
\hat L_{x,B}\equiv  \unit_A\otimes \hat L_x\, ,\label{lxAlxB}\\
\hat{L}_{x}&=&\hat L_{x,A}+\hat L_{x,B}\, , \label{lxcollective}
\eeqa
where $\unit_A$ and $\unit_B$ are unit operators in their respective
subspaces. Eq.(\ref{lxAlxB}) simply means that $\hat{L}_{x,A}$ acts
on the first (or ``A'') subsystem only, leaving the second (or
``B'') subsystem alone, and similarly for $\hat{L}_{x,B}$.  The
operators
\beqa
\hat{L}_{y} &=&\hat{L}_{y,A}+\hat{L}_{y,B},  \label{lycollective} \\
\hat{L}_{z} &=&\hat{L}_{z,A}+\hat{L}_{z,B}  \label{lzcollective}
\end{eqnarray}%
are defined in a similar obvious manner.

Let $\vert \chi ( \alpha ) \rangle _{A}$ and $\vert \phi ( \alpha )
\rangle _{B}$ be states of subsystems A and B, respectively, with
the property that
\begin{eqnarray}
( \hat{L}_{x,A}-i\alpha \hat{L}_{y,A}) \vert \chi ( \alpha ) \rangle
_{A}
&=&\lambda _{A}\vert \chi ( \alpha ) \rangle _{A} \\
( \hat{L}_{x,B}-i\alpha \hat{L}_{y,B} )\vert \phi ( \alpha ) \rangle
_{B} &=&\nu _{B}\vert \phi ( \alpha ) \rangle _{B}\, ,
\end{eqnarray}%
\emph{i.e.} $\left\vert \chi \left( \alpha \right) \right\rangle
_{A}$ and $\left\vert \phi \left( \alpha \right) \right\rangle _{B}$
are intelligent in their respective subsystems. Then,
\begin{equation}
\left\vert \psi (\alpha )\right\rangle =\left\vert \chi \left(
\alpha \right) \right\rangle _{A}\otimes \left\vert \phi \left(
\alpha \right) \right\rangle _{B}\equiv \left\vert \chi \left(
\alpha \right) \right\rangle _{A} \left\vert \phi \left( \alpha
\right) \right\rangle _{B}
\end{equation}%
is intelligent since
\begin{eqnarray}
( \hat{L}_{x}-i\alpha \hat{L}_{y}) \vert \psi ( \alpha ) \rangle
&=&\left[ ( \hat{L}_{x,A}-i\alpha \hat{L}_{y,A}) \vert \chi ( \alpha
) \rangle _{A}\right]
\vert \phi ( \alpha ) \rangle _{B}  \nonumber \\
&&\quad +\vert \chi ( \alpha ) \rangle _{A} \left[ ( \hat{L}_{x,B}-i\alpha \hat{L}%
_{y,B}) \vert \phi ( \alpha ) \rangle _{B}\right] \, , \\
&=& ( \lambda _{A}+\nu _{B}) \vert \chi ( \alpha ) \rangle _{A}
\vert \phi ( \alpha ) \rangle _{B}.  \label{compointelligent}
\end{eqnarray}%
In other words, the direct product of two intelligent states is also
intelligent, provided that one thinks of the resulting state as a
composite state constructed from two separate systems. This simple
result is quite powerful as it indicates that intelligent states can
be ``built-up" by putting together other intelligent states.

Quite clearly, the task now at hand is to find the simplest
intelligent states and use them as building blocks to construct more
complicated ones.

\section{Intelligent states with $\ell=\myfrac{1}{2}$}\label{spinhalfcase}


Consider the simplest realization of $\hat L_x-i\alpha\hat L_y$. Using basis
states $\vert + \rangle$ and $\vert - \rangle$, for which
\begin{equation}
\hat{L}_{z}\mapsto \frac{1}{2}\left( {\renewcommand{\arraystretch}{0.8}%
\begin{array}{cc}
1 & 0 \\
0 & -1%
\end{array}%
}\right) ,\ {\renewcommand{\arraystretch}{0.8}
\hat{L}_{x}\mapsto \frac{1}{2}\left(
\begin{array}{cc}
0 & 1 \\
1 & 0%
\end{array}%
\right) ,\ \hat{L}_{y}\mapsto \frac{1}{2}\left(
\begin{array}{cc}
0 & -i \\
i & 0%
\end{array}
\right)} ,\label{2by2angularmomenta}
\end{equation}%
we obtain the $2\times 2$ matrix%
\begin{equation}
{\renewcommand{\arraystretch}{0.8} \hat L_x-i\alpha\hat L_y\mapsto \frac{1}{2%
}\left(
\begin{array}{cc}
0 & 1-\alpha \\
1+\alpha & 0%
\end{array}%
\right) }.
\end{equation}%
The (unnormalized) eigenstates, which are by definition intelligent states,
are just
\begin{equation}
{\renewcommand{\arraystretch}{0.8} \left(
\begin{array}{c}
1 \\
\frac{1+\alpha }{\sqrt{1-\alpha ^{2}}}%
\end{array}%
\right) ,\left(
\begin{array}{c}
1 \\
-\frac{1+\alpha }{\sqrt{1-\alpha ^{2}}}%
\end{array}%
\right)}\, ,
\end{equation}
with respective eigenvalues
\beq
\lambda_+=\lambda\equiv\textstyle\frac{1}{2}\sqrt{1-\alpha^2}\,
,\qquad \lambda_-=-\lambda\, .\label{lambdadef}
\eeq
Introducing the quantity
\begin{equation}
\mu =\frac{1+\alpha }{\sqrt{1-\alpha ^{2}}},  \label{mudef}
\end{equation}
we obtain the normalized intelligent states as
\begin{eqnarray}
\vert \psi^{\myfrac{1}{2}}_{-}( \mu ) \rangle &=&
\frac{1}{\sqrt{1+\vert \mu\vert^{2}}} \left(
\begin{array}{c}
1 \\
-\mu%
\end{array}%
\right) =\frac{1}{\sqrt{1+\vert \mu \vert^{2}}}\,\vert + \rangle
-\frac{\mu
}{\sqrt{1+\vert \mu \vert ^{2}}}\,\vert - \rangle ,  \label{spinhalf1} \\
\vert \psi^{\myfrac{1}{2}} _{+}( \mu ) \rangle
&=&\frac{1}{\sqrt{1+\left\vert \mu \right\vert ^{2}}}\left(
\begin{array}{c}
1 \\
\mu%
\end{array}%
\right) =\frac{1}{\sqrt{1+\left\vert \mu \right\vert ^{2}}}\,\vert +
\rangle +\frac{\mu }{\sqrt{1+\left\vert \mu \right\vert
^{2}}}\,\vert - \rangle .  \label{spinhalf2}
\end{eqnarray}

We note that, if $\vert\alpha\vert <1$, $\mu$ is real and we can
write
\begin{equation}
\vert \psi^{\myfrac{1}{2}}_{\pm}(\beta) \rangle=R_y(\pm \beta)\vert
+ \rangle=\e^{\mp i\beta\,\hat L_y}\ket{+}\, , \label{rotated+}
\end{equation}
with
\begin{equation}
\cos\textstyle\frac{\beta}{2}=\frac{1}{\sqrt{1+\vert \mu\vert^2}}\, ,
\quad \sin\textstyle\frac{\beta}{2}=\frac{%
\mu}{\sqrt{1+\vert \mu\vert^2}}\, .\label{trigdefa+}
\end{equation}
\emph{From Eqn.(\ref{su2coherent}), we see that the
spin-$\myfrac{1}{2}$ intelligent states are also coherent states
when $\mu$ is real.}

On the other hand, when $\vert \alpha\vert \ge 1, \mu$ is purely
imaginary and we have
\begin{equation}
\vert \psi^{\myfrac{1}{2}}_{\pm}(\beta) \rangle=R_x(\pm \beta)\vert
+ \rangle=\e^{\mp i\beta \hat L_x}\ket{+}\, ,
\end{equation}
where, this time,
\begin{equation}
\cos\textstyle\frac{\beta}{2}=\frac{1}{\sqrt{1+\vert \mu\vert^2}}\, ,
\quad i\sin\textstyle\frac{\beta}{2}=\frac{%
\mu}{\sqrt{1+\vert \mu\vert^2}}\, .
\end{equation}
\emph{From Eqn.(\ref{su2coherent}), we see that the
spin-$\myfrac{1}{2}$ intelligent states are also coherent states
when $\mu$ is purely imaginary.}

\section{General construction}\label{generalcase}

\subsection{Example: An intelligent states with
$\ell=\myfrac{5}{2}$}\label{spin5halfcase}

We can use the states $\ket{\psi_{\pm}^{\myfrac{1}{2}}(\beta)}$ of
Eqns.(\ref{spinhalf1}) and (\ref{spinhalf2}) to construct
$\ell=\myfrac{5}{2}$ intelligent states as follows. Consider the
product
\beqa
\vert \psi_{+++--}(\beta) \rangle &=&\left[\vert
\psi^{\myfrac{1}{2}}_+(\beta) \rangle_1\,\vert
\psi^{\myfrac{1}{2}}_+(\beta) \rangle_2\,\vert
\psi^{\myfrac{1}{2}}_+(\beta) \rangle_3 \right]\nonumber \\
&&\qquad \otimes \left[\vert \psi^{\myfrac{1}{2}}_+(\beta)
\rangle_4\,\vert \psi^{\myfrac{1}{2}}_+(\beta) \rangle_5 \right]\, .
\eeqa
Here, the index $i$ labels one of five spin-$\hf$ subsystems. If we
expand every $\vert \psi^{\myfrac{1}{2}}_+(\beta) \rangle_i$ and
distribute the product, the first term of the resulting expression
is given by
\begin{equation}
\vert \ell=\myfrac{5}{2}\, ,m=\myfrac{5}{2} \rangle=\cos^5%
\textstyle\frac{\beta}{2} \left(\vert + \rangle_1\vert +
\rangle_2\vert + \rangle_3\vert + \rangle_4\vert +
\rangle_5\right)\, .
\end{equation}
This term is fully symmetric under permutation.

Let us use the shorthands
\begin{eqnarray}
\hat L_{x,1}&=&\hat L_{x}\otimes \mathinner{\hbox{1}\mkern-4mu\hbox{l}}_2%
\otimes\mathinner{\hbox{1}\mkern-4mu\hbox{l}}_3\otimes\mathinner{\hbox{1}%
\mkern-4mu\hbox{l}}_4\otimes \mathinner{\hbox{1}\mkern-4mu\hbox{l}}_5\, ,\nonumber  \\
\hat L_{x,2}&=& \mathinner{\hbox{1}\mkern-4mu\hbox{l}}_1\otimes
\hat L_{x}\otimes \mathinner{\hbox{1}\mkern-4mu\hbox{l}}_3\otimes%
\mathinner{\hbox{1}\mkern-4mu\hbox{l}}_4
\otimes\mathinner{\hbox{1}\mkern-4mu\hbox{l}}_5
\end{eqnarray}
\textit{etc.}, so that each $\hat L_{x,i}$ acts only on the $i$'th
subspace (of dimension 2).  Let
\beqa
\hat L_{x,A}&=&\hat L_{x,1}+\hat L_{x,2}+\hat L_{x,3}\, , \nonumber \\
\hat L_{x,B}&=&\hat L_{x,4}+\hat L_{x,5}\, ,
\eeqa
and define
\begin{equation} \hat L_x=\hat L_{x,A}+\hat L_{x,B}\, .
\label{collectiveAB}
\end{equation}
The collective operators $\hat L_y$ and $\hat L_z$ are defined similarly, as
are $\hat L_{\pm}$:
\begin{equation}
\hat L_{\pm}=\hat L_x\pm i\,\hat L_y\, .
\end{equation}

Because the collective operators are fully symmetric under
permutation of any two subspace index $i$ in
Eq.(\ref{collectiveAB}),
and act on the symmetric state $\vert \textstyle\frac{5}{2},\textstyle%
\frac{5}{2} \rangle$, every state of angular momentum
$\ell=\myfrac{5}{2}$ will be symmetric under permutation.  Thus, the
order in which the $\ket{+}$'s or $\ket{-}$'s occur is unimportant.

\subsubsection{The case $\vert \alpha\vert <1$}\label{secalphaless1}

With $\vert\alpha\vert <1$, every
$\ket{\psi_{\pm}^{\myfrac{1}{2}}(\beta)}$ is obtained by rotation
about the $y$--axis.  Thus, we can write
\begin{equation}
\vert \psi_{+++--}(\beta) \rangle=\left[R^A_y(\beta)\vert \textstyle\frac{3}{2},%
\textstyle\frac{3}{2} \rangle_A\right] \left[R^B_y(-\beta)\vert 1,1
\rangle_B \right]\, ,  \label{coupledstate}
\end{equation}
where we have directly coupled
\begin{eqnarray}
\left[R_y(\beta)\vert + \rangle_1\right]\otimes
\left[R_y(\beta)\vert + \rangle_2\right]\otimes
\left[R_y(\beta)\vert + \rangle_3\right] &=& R^A_y(\beta)
\left[\vert + \rangle_1\vert + \rangle_2\vert + \rangle_3\right]\, , \nonumber \\
&=&R^A_y(\beta)\vert \textstyle\frac{3}{2},\textstyle\frac{3}{2}
\rangle_A\, , \label{coherent3half}\\
\left[R_y(-\beta)\vert + \rangle_4\right]\otimes
\left[R_y(-\beta)\vert + \rangle_5\right]&=&R^B_y(-\beta)\vert 1,1
\rangle_B\, , \label{coherentone}
\end{eqnarray}
Here, the rotation operator $R^A_y(\beta)=\e^{-i\beta\hat L_{y,A}}$
while  $R^B_y(-\beta)=\e^{i\beta \hat L_{y,B}}$.  Note that the
states of Eqs.(\ref{coherent3half}) and (\ref{coherentone}) are both
angular momentum coherent states.

Eq.(\ref{coupledstate}) can now be expanded as
\begin{equation}
\displaystyle\sum_{m_A,m_B}\vert \3hf,m_A \rangle_A\,\vert 1,m_B
\rangle_B\,d^{\myfrac{3}{2}}_{m_A,\myfrac{3}{2}}(\beta)\,
d^{1}_{m_B,1}(-\beta)\, ,
\end{equation}
where
\beq
d^\ell_{m,m'}(\beta)\equiv\bra{\ell,m}\,R_y(\beta)\ket{\ell,m'}
\eeq
is the reduced Wigner function \cite{varshalovich}.

To project into the $\ell=\myfrac{5}{2}$ subspace, we specialize the
projector
\beq
\hat
\Pi^{\ell}=\displaystyle\sum_{m=-\ell}^{\ell}\ket{\ell,m}\bra{\ell,m}
\label{projectorell}
\eeq
to $\ell=\myfrac{5}{2}$ so as to obtain
\beq
\ket{\psi_{+++--}^{\myfrac{5}{2}}(\beta)
}\propto\displaystyle\sum_{m}\ket{\myfrac{5}{2},m}\,\kappa^{\myfrac{5}{2},m}_{\myfrac{3}{2},1}(\beta)\,
,\label{l52notnormalized}
\eeq
where
\beq
\kappa^{\ell m}_{\ell_A,\ell_B}(\beta)=\displaystyle\sum_{m_A (m_B)}
\cg{\ell_A}{m_A}{\ell_B}{m_B}{\ell}{m} \times
d^{\ell_A}_{m_A,\ell_A}(-\beta)\, d^{\ell_B}_{m_B,\ell_B}(\beta)\, ,
\label{kappacoeff}
\eeq
and $\cg{\ell_A}{m_A}{\ell_B}{m_B}{\ell}{m}$ is an su(2)
Clebsch-Gordan coefficient.

A better, more compact notation for $\vert
\psi_{+++--}^{\myfrac{5}{2}} \rangle $ is
\begin{equation}
\vert \psi_{+++--}^{\myfrac{5}{2}} \rangle\equiv \vert \psi_{\myfrac{3}{2},1}^{%
\myfrac{5}{2}}(\beta) \rangle\, . \label{l1l2notation}
\end{equation}
This emphasizes that only the total number of $\vert + \rangle_i$
states and the total number of $\vert - \rangle_j$ states are
relevant for the construction of an intelligent state of angular
momentum $\ell=\ell_A+\ell_B$.  The state
$\ket{\psi_{++-+-}^{\myfrac{5}{2}}(\beta)}$, for instance, can
differ from $\ket{\psi_{+++--}^{\myfrac{5}{2}}(\beta)}$ by at most a
phase.

To show that the state of Eq.(\ref{l1l2notation}) is intelligent, we
note that the operator $\hat\Pi^{\myfrac{5}{2}}$ of
Eq.(\ref{projectorell}) acts as the unit operator on any state
completely in the $\ell=\myfrac{5}{2}$ subspace, and annihilates any
state with no part in this subspace.  Hence, the collective $\hat
L_y=\hat L_{y,A}+\hat L_{y,B}$ operator and its $\hat L_x$
counterpart must commute with the projection
$\hat\Pi^{\myfrac{5}{2}}$ of Eq.(\ref{projectorell}) since neither
$\hat L_y$ nor $\hat L_x$ can change $\ell$. Thus,
\beqa
\left(\hat L_y-i\alpha\hat
L_x\right)\ket{\psi^{\myfrac{5}{2}}_{\myfrac{3}{2},1}(\beta)}
&=&\hat\Pi^{\myfrac{5}{2}}\left(\hat L_y-i\alpha\hat
L_x\right)\ket{\psi_{\myfrac{3}{2},1}(\beta)}\,
,\\
&=&(3\lambda_++2\lambda_-)\ket{\psi^{\myfrac{5}{2}}_{\myfrac{3}{2},1}(\beta)}\,
.
\eeqa
The projection does not preserve the norm so $\vert
\psi^{\myfrac{5}{2}}_{\myfrac{3}{2},1}(\beta) \rangle$ must be
normalized after the projection.

Since $R^A_y(\beta)\ket{\frac{3}{2},\frac{3}{2}}_A$ and
$R^B_y(-\beta)\ket{1,1}_B$ are coherent, we see that
$\ket{\psi^{\myfrac{5}{2}}_{\myfrac{3}{2},1}(\beta)}$ is the result
of coupling two su(2) coherent states.

\subsubsection{The case $\vert \alpha\vert \ge 1$.}

In this case, we note that
\begin{eqnarray}
&& \langle \ell,m \vert\,R_x(\beta)\vert \ell,\ell \rangle \nonumber
\\
&&\quad =\langle \ell,m \vert\,
R_z(-\myfrac{\pi}{2})\,R_y(\beta)\,R_z(\myfrac{\pi}{2})\,\vert \ell,\ell \rangle\, , \\
&&\quad=\hbox{\rm e}^{-i\pi(\ell-m)/2}d^{\ell}_{m,\ell}(\beta)\, ,
\end{eqnarray}
so that, for instance,
\begin{equation}
\vert \psi_{\myfrac{3}{2},1}^{\myfrac{5}{2}}(\beta) \rangle\propto
\displaystyle\sum_{m}\vert \textstyle\frac{5}{2},m
\rangle\,\hbox{\rm e}^{-i\pi(\frac{5}{2}-m)/2}
\,\kappa^{\myfrac{5}{2}m}_{\myfrac{3}{2},1}(\beta)\, ,
\end{equation}
is intelligent by the same argument given for the $\vert \alpha\vert
<1$ case.

\subsection{A general expression}\label{generalcase}

More generally, it is now clear that if we start with $2\ell_A$
copies of $\ket{\psi^{\myfrac{1}{2}}_+(\beta)}$ and $2\ell_B$ copies
of $\ket{\psi^{\myfrac{1}{2}}_-(\beta)}$, we can write
\beq
\left[R^A_y(\beta)\ket{\ell_A,\ell_A}\right]\otimes\left[R^B_y(-\beta)\ket{\ell_B,\ell_B}\right]\,
, \label{coupledexpression}
\eeq
and project into a good $\ell$ subspace using
Eq.(\ref{projectorell}) to obtain an intelligent state of angular
momentum $ \ell=\ell_A+\ell_B$ as
\begin{equation}
\vert \psi^{\ell}_{\ell_A,\ell_B}(\beta) \rangle\propto\displaystyle%
\sum_{m}\vert \ell,m
\rangle\,\kappa_{\ell_A,\ell_B}^{\ell,m}(\beta)\, ,
\label{finalexpression}
\end{equation}
with $\kappa_{\ell_A,\ell_B}^{\ell,m}(\beta)$ given in
Eq.(\ref{kappacoeff}).

Eqs.(\ref{coupledexpression}) and (\ref{finalexpression}) show
explicitly how su(2) intelligent states with angular momentum $\ell$
can be constructed by appropriately coupling su(2) coherent states.
The state of Eq.(\ref{coupledexpression}) is explicitly intelligent
and remains intelligent under projection by $\hat\Pi^{\ell}$ of
Eq.(\ref{projectorell}), thus yielding Eq.(\ref{finalexpression}).

We show in \ref{coefficient} how
$\kappa_{\ell_A,\ell_B}^{\ell,m}(\beta)$ can be reduced to
\beq
\kappa_{\ell_A\ell_B}^{\ell ,m}(\beta)= 2^{\ell }\frac{\sqrt{\left(
2\ell_B\right) !\left( 2\ell_A\right) !\left( \ell +m\right) !\left(
\ell -m\right) !}}{\left( 2\ell \right) !} \,
d_{\ell_B-\ell_A,m}^{\ell }\left( \textstyle\frac{\pi
}{2}\right)\,d^{\ell}_{m,\ell}(\beta) . \label{cfinal}
\eeq

Introducing the norm
\beq
{\cal
N}^{\ell}_{\ell_A,\ell_B}(\beta)=\frac{1}{\sqrt{\displaystyle\sum_{m}\vert
\kappa^{\ell,m}_{\ell_A,\ell_B}(\beta)\vert^2}} \, ,
\eeq
we obtain the final expression for our intelligent state as
\beq
\ket{\psi^{\ell}_{\ell_A,\ell_B}(\beta)}={\cal
N}^{\ell}_{\ell_A,\ell_B}(\beta)\,\displaystyle\sum_{m}
\ket{\ell,m}\, \kappa^{\ell,m}_{\ell_A,\ell_B}(\beta)\, .
\label{generalstate}
\eeq

Finally, we note that the eigenvalue problem in the
$\ell=\ell_A+\ell_B$ subspace has at most $2\ell+1$ independent
eigenvectors. Using Eq.(\ref{generalstate}), it is clear that
(except when $\beta=0$ or $\pi$) we can construct exactly the right
number linearly independent states of the form by selecting in turn
$(\ell_A,\ell_B)$ to be $(\ell,0),(\ell-
\myfrac{1}{2},\myfrac{1}{2}),\ldots,(0,\ell)$. Hence, all $2\ell+1$
intelligent states are coupled su(2) coherent states.

When $\beta=0$ or $\pi$, $\alpha$ is $\mp 1$, the operator $\hat
L_x-i\alpha \hat L_y$ is nilpotent and has a single eigenvector:
either $\ket{\ell,\ell}$ or $\ket{\ell,-\ell}$.  This is
(indirectly) illustrated in Figures \ref{Figj5hf} and \ref{Figj3},
where it is that all uncertainty curves merge to a single curve at
$\beta=\pi$ (or $\alpha=1$).

\section{Selected results}\label{results}

\subsection{Expectations and standard deviations}

The intelligent state of Eq.(\ref{generalstate}) is an eigenstate of
$\hat L_x-i\alpha \hat L_y$ with eigenvalue
\beq
\lambda_{\ell_A,\ell_B}=\lambda\,(2\ell_A-2\ell_B)\, .
\eeq

If we assume $\vert\alpha\vert \le 1$, then $\lambda$ is real.
Combining Eqs.(\ref{lambdadef}),(\ref{mudef}) and (\ref{trigdefa+}),
we obtain
\beq
\lambda_{\ell_A,\ell_B}=(\ell_A-\ell_B)\sin\beta\, .
\eeq
Since $\alpha,\langle\,\hat L_x\rangle$ and $\langle\,\hat
L_y\rangle$ are real, this can be compared with
$\lambda_{\ell_A,\ell_B}=\langle\,\hat L_x\rangle-i\alpha
\langle\,\hat L_y\rangle$ to give
\beq
\langle\,\hat L_x\rangle=\hf\left(\ell_B-\ell_A\right)\sin\beta\,
,\qquad\langle\,\hat L_y\rangle=0\, .
\eeq

If, on the other hand, $\vert \alpha \vert \ge 1$, we have
\beq
\langle\,\hat L_y\rangle=-\hf\left(\ell_B-\ell_A\right)\sin\beta\,
,\qquad\langle\,\hat L_x\rangle=0\, .
\eeq

Furthermore, using Eqs.(\ref{collinear}) and (\ref{anticondition}),
one finds that the intelligent states generally
satisfy
\beq
\left(\Delta L_y\right)^2=-\frac{1}{2\alpha}\,\langle\,\hat
L_z\rangle\ ,\qquad  \left(\Delta
L_x\right)^2=-\frac{1}{2}\alpha\,\langle\,\hat L_z\rangle\, .
\label{deltalydeltalx}
\eeq
This allows computation of all pertinent quantities from
$\langle\,\hat L_z\rangle$, which is simply given by
\beq
\langle\,\hat
L_z\rangle=\left(\mathcal{N}_{\ell_A\ell_B}^{\ell}(\beta)\right)^2\,
\left(\displaystyle\sum_m\,m\,\vert\,\kappa^{\ell
m}_{\ell_A,\ell_B}(\beta)\,\vert^2\right)\, . \label{avglzI}
\eeq

\subsection{Numerical results}

\begin{figure}[htp]
\begin{center}
\quad\epsfysize= 2.75 in \epsfbox{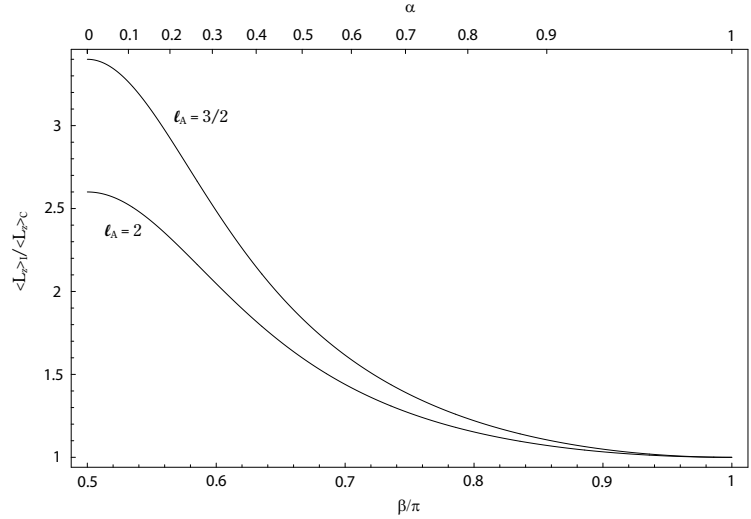}
\end{center}
\caption{The ratio $\vert\,\langle\,\hat
L_z\rangle\,\vert_I/\vert\,\langle\,\hat L_z\rangle\,\vert_c$ as a
function of $\myfrac{\beta}{\pi}$ or $\alpha$ for
$\ell=\myfrac{5}{2}$ and various values of $\ell_A$ and $\ell_B$ so
that $\ell_A+\ell_B=\myfrac{5}{2}$.} \label{Figj5hf}
\end{figure}

\begin{figure}[htp]
\begin{center}
\quad\epsfysize= 2.75 in \epsfbox{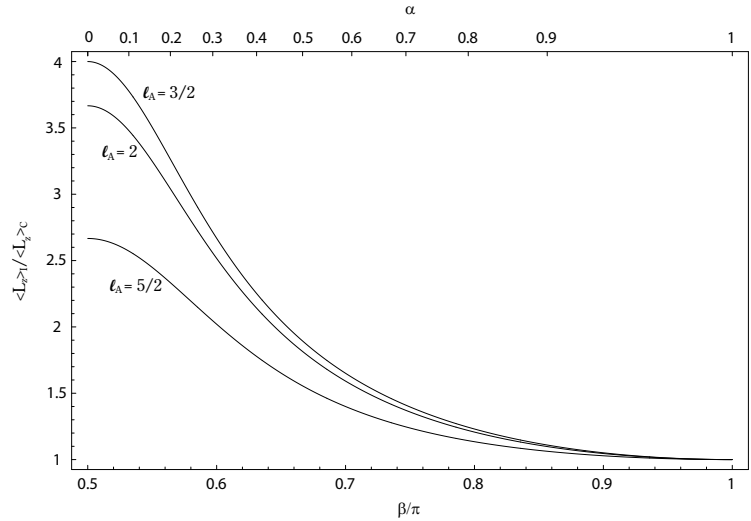}
\end{center}
\caption{The ratio $\vert\,\langle\,\hat
L_z\rangle\,\vert_I/\vert\,\langle\,\hat L_z\rangle\,\vert_c$ as a
function of $\myfrac{\beta}{\pi}$ or $\alpha$ for $\ell=3$ and
various values of $\ell_A$ and $\ell_B$ so that $\ell_A+\ell_B=3$.}
\label{Figj3}
\end{figure}

Figures \ref{Figj5hf} and \ref{Figj3} illustrate typical results.
The figures give the ratio of the uncertainty products $(\Delta
L_x\,\Delta L_y)_I$ of intelligent states to the coherent state
$(\Delta L_x\,\Delta L_y)_c$, for which $\ell_A=\ell$.  These ratios
are just the ratios of $\langle  \hat  L_z\rangle$.  For the
coherent state, one rapidly finds
\beq
\langle\,\hat L_z\,\rangle_c=\frac{\ell}{2}\cos\beta\, ,
\eeq
for $\vert\alpha\vert<1$.

In figure \ref{Figj5hf}, the ratios for intelligent states of
angular momentum $\ell=\myfrac{5}{2}$ with
($\ell_A=2,\ell_B=\myfrac{1}{2}$) and ($\ell_A=\myfrac{3}{2},
\ell_B=1$) are given.  The results are unchanged if one switches
$\ell_A$ and $\ell_B$.  The curves $\alpha <0$ are identical to
those for $\alpha
>0$.
 Furthermore, the results with
$\vert \alpha\vert >1$ can be obtained from those with
$\vert\alpha\vert<1$ by the transformation
$\alpha\to -\myfrac{1}{\alpha}$, so the range $0\le \alpha
\le 1$ captures all qualitative features of the curves. Figure
\ref{Figj3} is similar to \ref{Figj5hf}, except that
$\ell=3$.  The symmetries of Fig.\ref{Figj5hf} are also present in
Fig. \ref{Figj3}.

One immediately observes that the uncertainty products for
intelligent states (with $\ell_A\ne \ell$) is always greater than
the corresponding product for the coherent state (with
$\ell_A=\ell$).  Insofar as the product $\Delta L_x\,\Delta L_y$
goes, the ``worst" intelligent state is the one for which $\ell_A$
and $\ell_B$ are as close as possible.  We have not been able to
prove this analytically because the expression (\ref{avglzI}) for
$\langle\,\hat L_z\,\rangle$ is difficult to manipulate.  However,
we have verified that this observation holds over a wide range of
values of $\ell$.  Other curves illustrating this behavior can be
found in \cite{polynomialstates}.

It is not difficult to show that the maximum of the product $\Delta
L_x\Delta L_y$ is simply $\hf\ell$.  Indeed, by
Eq.(\ref{intelligentdef}), it is clear that the product is maximal
when $\vert \langle\hat L_z\rangle\vert$ is maximal.  This maximum
is reached for the states $\ket{\ell,\pm\ell}$. From
Eqs.(\ref{coupledexpression}) and (\ref{finalexpression}), it
immediately follows that this will occur when $\beta=0$ or
$\beta=\pi$.  This, implies by Eq.(\ref{trigdefa+}) that $\mu=0$ or
$\mu=\infty$ which in turn,by Eq.(\ref{mudef}), implies
$\alpha=\pm\,1$.

As $\alpha\to\pm 1$, all intelligent states converge to a single
state.  When $\alpha=\pm 1$ precisely, the operator $\hat
L_x-i\alpha\hat L_y$ becomes the nilpotent $\hat L_+$ or $\hat L_-$
respectively, both of which have only one non-zero eigenvector.

\begin{figure}[htp]
\begin{center}
\quad\epsfysize= 2.75 in \epsfbox{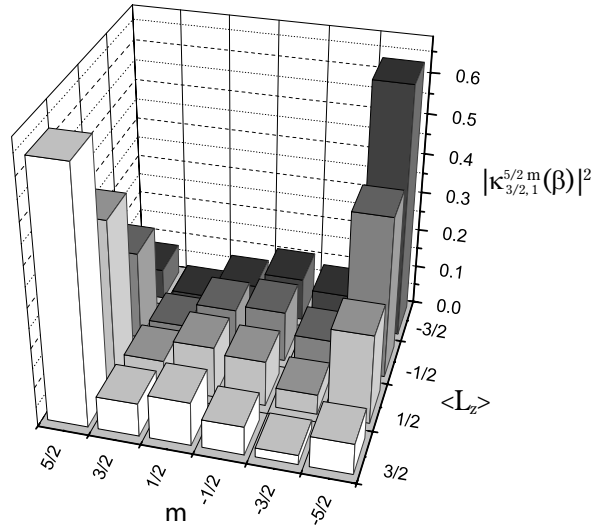}
\end{center}
\caption{The populations of $m$ substates $\vert
\kappa^{\myfrac{5}{2}\, m}_{\myfrac{3}{2},1}(\beta)\vert^2$ for
different values of $m$ and $\ell=\myfrac{5}{2}$. The values of
$\beta$ were selected so that $\langle\,\hat L_z\rangle=
\pm\myfrac{3}{2},\pm\myfrac{1}{2}$.} \label{FigCjm}
\end{figure}

Figure \ref{FigCjm} shows the population of various $m$ substates in
the intelligent state
$\ket{\psi^{\myfrac{5}{2}}_{\myfrac{3}{2},1}(\beta)}$.  For clarity,
we have restricted the calculations to angles $\beta$ chosen so that
$\langle\,\hat L_z\rangle= \pm\myfrac{3}{2},\pm\myfrac{1}{2}$.

This figure illustrates a very general symmetry: $\vert
\kappa^{\ell,m}_{\ell_A,\ell_B}(\beta)\vert^2= \vert
\kappa^{\ell,-m}_{\ell_A,\ell_B}(-\beta)\vert^2$. This can be traced
back to symmetries of the $d$-functions entering in the construction
of the $\kappa^{\ell,m}_{\ell_A,\ell_B}(\beta)$ coefficients.

\section{Discussion and conclusion}\label{discussion}

Let us construct the operators
\begin{equation}
\hat K_{i}=\hat L_{iA}-\hat L_{iB}.
\end{equation}%
The operators $\{ \hat K_{x},\hat K_{y},\hat K_{z}\} $ do not close
under commutation.  However, ${ \hat K_{x},\hat K_{y},\hat L_{z}} $
do close on an angular momentum algebra, which we call the $K$-
angular momentum $su(2)_K$.  This set is interesting because our
intelligent states are constructed using a $K$- rotation about $y.$
Indeed, defining $\hat K_{\pm }$ in the usual manner, one can see
that the state
\beq
\ket{\ell_A,\ell_A}\ket{\ell_B,\ell_B}\, ,\label{coupledextremal}
\eeq
is an eigenstate of $\hat L_{z}$ with eigenvalue $m_{K}=\ell
_{A}+\ell _{B}=\ell$. Because (\ref{coupledextremal}) is killed by
$\hat K_{+},$ it can be identified with the state $\left\vert \ell
,m_{K}=\ell \right\rangle _{K}$ of $K$-angular momentum. In
particular, our starting state
\begin{eqnarray}
\left[ R_{y}^{A}(\beta )\left\vert \ell _{A},\ell _{A}\right\rangle \right] %
\left[ R_{y}^{B}(-\beta )\left\vert \ell _{B},\ell _{B}\right\rangle
\right]&=&\exp \left[ -i\beta \left( \hat L_{yA}-\hat L_{yB}\right)
\right] \left\vert \ell
,\ell \right\rangle _{K} \nonumber \\
&=&\exp \left[ -i\beta \hat K_{y}\right] \left\vert \ell ,\ell
\right\rangle_K
\end{eqnarray}%
and is thus a $K$-angular momentum coherent state.

Unfortunately, the $K$-angular momenta do not commute with the
collective angular momenta $\hat L_{i}.$ Although
(\ref{coupledextremal}) is simultaneously a state with ``good" total
$\ell ,m_{\ell }$ and ``good" $\ell _{K}=\ell\, ,m_{K},$ other
$\left\vert \ell ,m_{K}\neq \ell \right\rangle _{K}$ states
generated by the action of $\hat K_{-}$ do not have ``good" $\ell
,m_{\ell };$ hence the need for the projection into the subspace of
good $L$- angular momentum.


In \cite{polynomialstates}, a method of constructing all intelligent
states of angular momentum $\ell$ was proposed.  The basic
polynomials
$\xi^x$ and $\eta^y$ are related to the direct product of
$x$ copies of
$\ket{+}$ and the direct product of $y$ copies
$\ket{-}$, respectively, via the correspondences
\beq
\ket{+}\leftrightarrow\xi\, ,\quad \ket{-}\leftrightarrow\eta\
,\quad
\ket{\ell,m}\leftrightarrow\frac{\xi^{\ell+m}\eta^{\ell-m}}{\sqrt{(\ell+m)!(\ell-m)!}}\,
.
\eeq
Using this, we can write, for $\vert \alpha\vert < 1$, the
intelligent state $\ket{\psi^{\ell}_{\ell_A,\ell_B}(\beta)}$ as the
product
\beq
\ket{\psi^{\ell}_{\ell_A,\ell_B}(\beta)}
=\left(\xi\,\cos\textstyle\frac{\beta}{2}+\eta\,\sin\frac{\beta}{2}\right)^{2\ell_A}\,
\left(\xi\,\cos\textstyle\frac{\beta}{2}-\eta\,\sin\frac{\beta}{2}\right)^{2\ell_B}\,
.\label{polystates}
\eeq
There is no need for projection as the result is a polynomial of
total degree $2\ell=2(\ell_A+\ell_B)$.  It is well--known that the
polynomials of the form $\xi^x\eta^y$, with $x+y=2\ell$, span a
basis for the su(2) representation of angular momentum $\ell$. The
combinatorics involved in the expansion of Eq.(\ref{polystates}) and
the conversion of various $\xi^x\eta^y$ to angular momentum states
yield precisely Eq.(\ref{generalstate}). Thus, we recover in a much
more transparent way the construction and calculations of
\cite{polynomialstates}.  (An expression similar to
Eq.(\ref{polystates}) can easily be found for $\vert \alpha\vert \ge
1$.)

The simple form of Eqs.(\ref{generalstate}),(\ref{deltalydeltalx})
and (\ref{avglzI}) illustrate the economy inherent to an approach
based on coupling.  These results can be contrasted, for instance,
with the corresponding expressions of \cite{rashid} or the
application done by \cite{Arvieu} of su(2) intelligent states in
nuclear physics.

Our results, which only require a table to Clebsch-Gordan
coefficient and expressions for Wigner $D$-function, represent the
simplest example of what could be a systematic algorithm for the
construction of intelligent states of observables elements of other
Lie algebras \cite{ben} or even deformed algebras \cite{burgos}. In
other words, the procedure presented here is easily generalizable.
Indeed, using the results of \cite{Sharp}\cite{su3paper}, the
properties of some SU(3) intelligent states will be the topic of a
forthcoming paper\cite{ben}.

\section{Acknowledgments}

This works is supported in part by NSERC of Canada, the Government
of Ontario and Lakehead University.  We would like to thank L. L.
S\'anchez-Soto and A. B. Klimov for valuable discussions, and S.
Buhmann for bringing to our attention some references used in the
preparation of this work.  HdG would like to thank A. Ballesteros
and F. Herranz for interesting suggestions on some aspects of the
current and future work.

\appendix

\section{The coefficient
$\kappa^{\ell,m}_{\ell_A,\ell_B}(\beta)$}\label{coefficient}

The expression of Eq.(\ref{cfinal}) can be manipulated into a more
transparent form using
\cite{varshalovich}
\beqa
d^{\ell_B}_{m_B,\ell_B}(-\beta)&=&(-1)^{m_B-\ell_B}\,d^{\ell_B}_{m_B,\ell_B}(%
\beta)\, ,\\
d^{\ell_A}_{m_A,\ell_A}(\beta)\,d^{\ell_B}_{m_B,\ell_B}(%
\beta)&=& \cg{\ell_A}{m_A}{\ell_B}{m_B}{\ell}{m}\times
d^{\ell}_{m,\ell}(\beta)\, ,
\eeqa
where $\ell=\ell_A+\ell_B$ and $\cg{\ell_A
}{\ell_A}{\ell_B}{\ell_B}{\ell}{\ell} =1$ have been used. Thus,
\beq
\kappa_{\ell_A,\ell_B}^{\ell,m}(\beta)=d^{\ell}_{m,\ell}(\beta)\times\left[
\displaystyle\sum_{m_A(m_B)}
(-1)^{m_B-\ell_B}\,\left(\cg{\ell_A}{m_A}{\ell_B}{m_B}{\ell}{m}\right)
^2\right] \, .
\eeq
A little more mileage can be done because Clebsch-Gordan
coefficients for which $\ell=\ell_A+\ell_B$ have known expressions
\cite{varshalovich}. Using this and the condition $m=m_A+m_B$, we
obtain
\begin{equation}
\kappa_{\ell_A,\ell_B}^{\ell,m}(\beta) =\frac{d^{\ell}_{m,\ell}(\beta)}{{\binom{{%
2\ell}}{{\ell-m}}}}\, \left[ \displaystyle\sum_{n=0}^{2\ell_B} (-1)^{n}\, {%
\binom{{2\ell_A}}{{\ell-m-n}}}{\binom{{2\ell_B}}{{n}}}\right]\, .
\end{equation}
The coefficient in the bracket can be identified with the
coefficient of $x^{\ell-m}$ in the expansion of
$(1+x)^{2\ell_A}(1-x)^{2\ell_B}$. In particular, when
$\ell_A=\ell_B$, $\ell$ is integer and there can be no odd powers of
$x$, so that no odd values of $m$ will appear in the expansion.

Finally \cite{varshalovich},
\beq
\sum_{n=0}^{2\ell_B}\left( -1\right) ^{n}\binom{2\ell_A}{\ell -m-n}%
\binom{2\ell_B}{n} =2^{\ell }\sqrt{\frac{%
\left( 2\ell_B\right) !\left( 2\ell_A\right) !}{\left( \ell
+m\right) !\left( \ell -m\right) !}}\,d_{\ell_B-\ell_A,m}^{\ell
}\left( \textstyle\frac{\pi }{2}\right) .
\eeq
Inserting this into $\kappa_{\ell_A\ell_B}^{\ell ,m}\left( \beta
\right)$ gives Eq.(\ref{cfinal}).

Note that the appearance of a rotation by $\myfrac{\pi}{2}$ about
the $\hat y$ axis:
\beqa
d_{m,\ell_B-\ell_A}^{\ell }\left(\textstyle\frac{\pi
}{2}\right)&=&d_{\ell_B-\ell_A,m}^{\ell }\left( -\textstyle\frac{\pi
}{2}\right)\nonumber
\\
&=&\bra{\ell,\ell_B-\ell_A}\e^{-i\frac{\pi}{2}\,\hat
L_y}\ket{\ell,m}\, ,
\eeqa
is reminiscent of an expression found in \cite{rashid}.

Lastly, although we have limited ourselves to expressions where
$\ell=\ell_A+\ell_B$, the factor $\ell_B-\ell_A$ makes it clear
that, up to a normalization, it is only the difference between
angular momenta that is here relevant.  More precisely, if one
considers $\ell_A^\prime=\ell_A+j, \ell_B^\prime=\ell_B+j$, then the
tensor product $\ell_A^{\prime}\otimes \ell_B^{\prime}$ will contain
a subspace of angular momentum $\ell$.  The coupled states in this
subspace are also intelligent, but are simply proportional to the
state obtained by coupling $\ell_A\otimes\ell_B$.  In other words,
no new state is found by considering cases other than
$\ell=\ell_A+\ell_B$.


\bigskip

\begin{thebibliography}{99}
\bibitem{robertson} H.P. Robertson,  Phys. Rev. \textbf{34}, 163-164
(1929)

\bibitem{aragone} C. Aragone \textit{et al.}, J. Phys. A: Math., Nucl. Gen.
\textbf{7}, L149-L151 (1974), C. Aragone \textit{et al.}, J. Math.
Phys. \textbf{17} 1963-1971 (1975)

\bibitem{FrankeArnold} Sonja Franke--Arnold \emph{et al}, New J. Phys. \textbf{6} 103
(2004),

\bibitem{Pegg} D T Pegg \emph{et al.}, New J. Phys. \textbf{7}, 62.1-62.20
(2005)

\bibitem{hradil} Z. Hradil \textit{et al}, arXiv:quant-ph/0605137

\bibitem{su2coherent} J. M. Radcliffe, J. Phys. A: Gen. Phys. \textbf{4}
313-323 (1971), F.T. Arecchi \textit{et al.} Phys. Rev. \textbf{A6}
2211-2237 (1972), R.R. Puri, \emph{Mathematical methods of Quantum Optics},
Springer--Verlag, Berlin Heidelberg 2001


\bibitem{Hillery} M. Hillery and L. Mlodinow, Phys. Rev. \textbf{A48}
1548-1558 (1993)

\bibitem{rashid} M. A. Rashid, J. Math. Phys. \textbf{19} 1391-1396 (1976)

\bibitem{opticalimplementation} B. Yurke, S. L. McCall and J. R. Klauder,
Phys. Rev. \textbf{A33} 4033-4054 (1986); M. Reck \emph{et al.},
Phys. Rev. Lett. \textbf{73} 58-61 (1994),

\bibitem{polynomialstates} Matthew M. Milks and Hubert de Guise,
J.Opt. B: Quantum SemiClass. Opt.\textbf{7} (2006) S622--S627,

\bibitem{Nha} Hyunchul Nha and Jaewan Kim, Phys. Rev. \textbf{A74}
012317 (2006),


\bibitem{brif1} C. Brif, Int. J. Theor. Phys. \textbf{36}, 1651-1682 (1997)

\bibitem{brif2} C. Brif and A. Mann, Quant. and Semiclass. Optics \textbf{9}
899-920 (1997), C. Brif and A. Mann, Phys. Rev. \textbf{A54} 4505-4518
(1996),

\bibitem{campos} R. A. Campos and C. C. Gerry, Phys. Rev. \textbf{A60}
1572-1576 (1999)

\bibitem{perinova} V. Pe\u{r}inov\'a, A Luk\u{s} and J. K\u{r}epelka, J.
Opt. B: Quantum Semiclass. Opt. \textbf{2} 81-89 (2000),

\

\bibitem{agarwalpuriI} G. S. Agarwal and R. R. Puri, Phys. Rev. \textbf{A49}
4968-4971 (1994)

\bibitem{trifonov} D. A. Trifonov, J.Phys. A: Math. Gen. \textbf{30}
5941-5957 (1997)

\bibitem{eigenvalueequation} W. H. Louisell, \emph{Quantum Statistical
Properties of Radiation}, Wiley-Interscience, N.Y. (1990); Asher Peres,
\emph{Quantum Theory: Concepts and Methods}, Kluwer Academic, Dordrecht, 1995

\bibitem{varshalovich} D. A. Varshalovich, A. N. Moskalev, V. K.
Khersonskii, \emph{Quantum Theory of Angular Momentum}, World
Scientifice, Singapore, (1988)

\bibitem{ben} Benjamin R. Lavoie and Hubert de Guise, \emph{SU(3) intelligent
states}, in preparation,

\bibitem{burgos} F.J. Herranz and A. Ballesteros, private
communication,

\bibitem{Arvieu} R. Arvieu and P. Rozmej, J.Phys. A: Math. Gen
\textbf{32} (1999) 2645--2652; P. Rozmej and R. Arvieu, Phys. Rev.
\textbf{A58} (1998) 4314--



\bibitem{Sharp} R. T. Sharp and Hans Von Baeyer, J. Math. Phys.
\textbf{7} (1966) 1105--1122; D. J. Rowe and C. Bahri, J. Math. Phys
\textbf{41} (2000) 6544--6565;

\bibitem{su3paper} D. J. Rowe, B. C. Sanders and H. de Guise,
J. Math. Phys. \textbf{40} (1999) 3604--3615


\end{thebibliography}
\end{document}